\documentclass[12pt]{iopart}
\usepackage{mathenv}
\usepackage[english]{babel}
\usepackage{mathabx}
\usepackage{graphicx,epstopdf}
\usepackage{bm}
\usepackage{cite}

\usepackage{color}
\usepackage[bookmarks,colorlinks,linkcolor=blue,anchorcolor=black,urlcolor=blue,citecolor=blue]{hyperref}
\usepackage{CJK}
\usepackage{esint}

\begin{document}
\begin{CJK*}{UTF8}{gbsn}

\title[Nonreciprocal recovery of electromagnetically...]{Nonreciprocal recovery of electromagnetically induced transparency by wavenumber mismatch in hot atoms}

\author{Lida Zhang~(\CJKfamily{gbsn}张理达)$^{1}$, Nina Stiesdal$^{2}$, Hannes Busche$^{2}$, Mikkel Gaard Hansen$^{2}$, Thomas Pohl$^{3}$, and Sebastian Hofferberth$^{2}$}
\address{$^{1}$School of Physics, East China University of Science and Technology, Shanghai, 200237, China}
\address{$^{2}$Institut f\"{u}r Angewandte Physik, University of Bonn, Wegelerstr. 8, 53115 Bonn, Germany}
\address{$^{3}$Institute for Theoretical Physics, Vienna University of Technology (TU Wien), 1040 Vienna, Austria}
\date{\today}

\begin{abstract}
For multi-level systems in hot atomic vapors the interplay between the Doppler shift due to atom velocity and the wavenubmer mismatch between driving laser fields strongly influences transmission and absorption properties of the atomic medium. In a three-level atomic ladder-system, Doppler broadening limits the visibility of electromagnetically-induced transparency (EIT) when the probe and control fields are co-propagating, while EIT is recovered under the opposite condition of counter-propagating geometry and $k_{p} < k_{c}$, with $k_{p}$ and $k_{c}$ being the wavenumbers of the probe and control fields, respectively. This effect has been studied and experimentally demonstrated as an efficient mechanism to realize non-reciprocal probe light transmission, opening promising avenues for example for realization of magnetic-field free optical isolators. 
In this tutorial we discuss the theoretical derivation of this effect and show the underlying mechanism to be an avoided crossing of the states dressed by the coupling laser as a function of atomic velocities when $k_{p}<k_{c}$. We investigate how the non-reciprocity scales with wavelength mismatch and show how to experimentally demonstrate the effect in a simple Rydberg-EIT system using thermal Rubidium atoms.
\end{abstract}

\maketitle

\end{CJK*}

\section{Introduction}

The phenomenon of electromagnetically induced transparency (EIT) has emerged as a key technique for quantum optics since the first observations three decades ago ~\cite{Harris1990PRL,harris1991,Fleischhauer2000PRL,Fleischhauer2002PRA,hau2001,lukin2001,marangos2005, sorensen2007}, allowing for example coherent photon storage and memory in atomic media \cite{Lukin2003,Adams2013,Kuzmich2013PRA_storage,Kuzmich2018,Firstenberg2018natcomm,Treutlein2022}. Introducing long-lived Rydberg-states into EIT ladder-schemes \cite{adams2007} enables novel applications such as electric and magnetic field sensing \cite{Adams2008,Shaffer2012b,Shaffer2013,Raithel2017,Weatherill2020,Cox2021,Shi2022b} and few-photon nonlinearities~\cite{adams2010,Saffman2010RMP,lukin2011, vuletic2012}. The latter have been used for single-photon sources~\cite{kuzmich2012,Pfau2018,Firstenberg2018,Polzik2021}, photon-atom entanglement \cite{Kuzmich2013}, single-photon switches~\cite{durr2014,rempe2014,hofferberth2014}, photon subtraction~\cite{hofferberth2016,hofferberth2021} and photonic quantum gates~\cite{durr2016,pohl2017,durr2019}. 

In hot atomic vapors, atomic motion and the resulting Doppler shift strongly influences the EIT effect\cite{xiao1995jan,Xiao1995PRL,dunn1996,Agarwal1996,Adams2001,adams2007,Moon2012,low2013}. More specifically, the observed transmission depends on the projection of the vector sum of the probe and control wavevectors $\mathbf{k}_{p}$ and $\mathbf{k}_{c}$ onto the probe beam direction. Interestingly, this makes three-level EIT systems inherently non-reciprocal in the sense that reversing the probe beam direction along the same propagation axis, while keeping the control field fixed, can result in completely different transmission \cite{Dunn1995PRA,dunn1999}.

For typical atomic lambda-systems with probe and control fields of approximately equal wavelength ($k_{p}=|\mathbf{k}_{p}| \approx k_{c}=|\mathbf{k}_{c}|$) the Doppler effect is mitigated by a co-propagating beam geometry \cite{Xiao1995april,tabosa2004}. However, the residual Doppler broadening due to the small wavenumber mismatch between the two fields leads to both a narrowing of the EIT window and nonvanishing absorption inside the window~\cite{xiao2012}. 
In contrast, in a three-level ladder-system with co-linear probe and control fields, where usually $k_{p} \not \approx k_{c}$, EIT can be greatly enhanced as well as attenuated depending on the ratio $k_c/k_p$ and the relative signs of the wavevectors $\mathbf{k}_c, \mathbf{k}_p$ \cite{dunn1999}. This allows the realization of nonreciprocal optics in vapor cells by correct choice of the atomic transitions used for EIT \cite{nori2021}. 

In this tutorial we discuss the interplay between EIT enhancement and attenuation and wavenumber mismatch in a ladder-type three-level atomic system. We show that for counter-propagating beams a gap appears in the eigenenergy spectrum of the two states dressed by the control field as a function of atomic velocity. Because probe photons cannot be absorbed within this avoided crossing, a box-shaped transmission window inside the Doppler-broadened absorption background opens, which width can be even larger than the EIT window in the case of cold (stationary) atoms under the same control field intensity \cite{dunn1999}. We study the dependence of the hot-vapor transmission window width and height on the wavenumber mismatch in detail and determine the optimal relation between $k_c$ and $k_p$ to maximize EIT width. While this ratio is not available for ground-state transitions in the alkali species commonly used in vapor cell EIT experiments \cite{Adams2001}, it can be achieved in these species in ladder configurations with Rydberg states \cite{adams2007}. We demonstrate the non-reciprocity in an experiment in hot Rubidium vapor, using $|20S\rangle$ as upper state in the ladder-system, and show that the experimental data agrees very well with the theory model. 

\begin{figure}[t]
 \centering
 \includegraphics{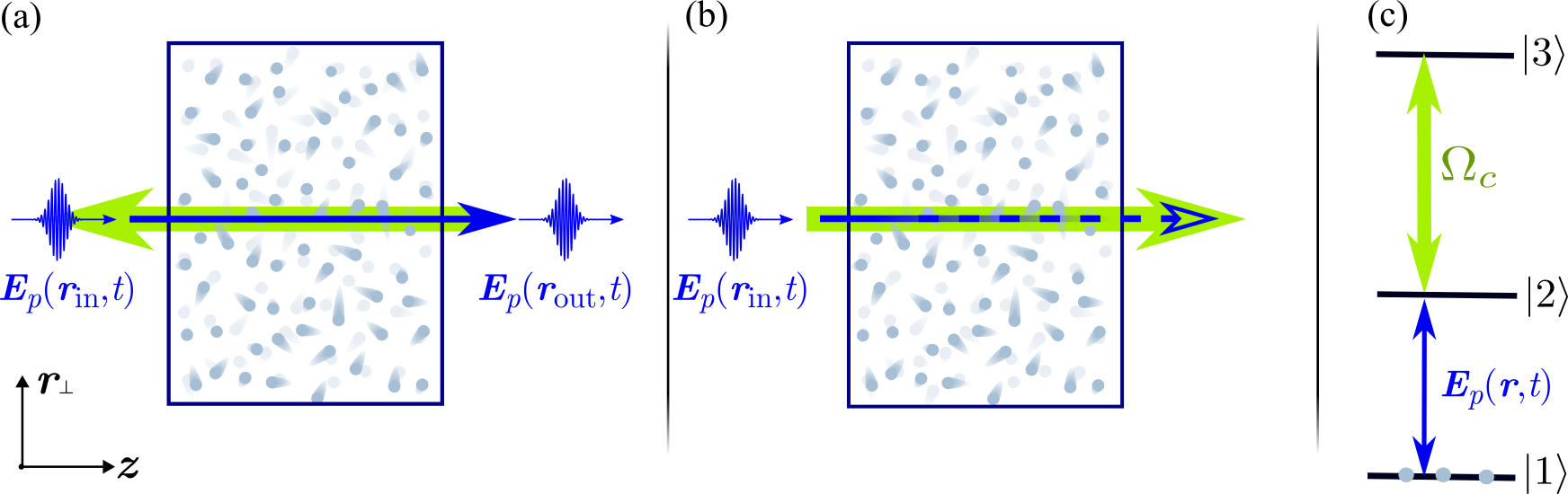} 
 \caption{(Color online) The probe $\bm{E}_{p}(\bm{r},t)$ and a classical laser field $\Omega_{c}$ are coupled to a three-level ladder-type hot atomic system shown in (c). When the two fields are counter-propagating, the EIT spectrum for the probe can be recovered in the condition of negative wavenumber mismatch such that the probe photon can fully pass through. When they are co-propagating, the EIT medium is turned into a broadband absorber such that the probe photon is randomly scattering into all directions due to strong absorption.}
 \label{fig:3-level-ladder}
\end{figure}

\section{\bf Theoretical model}
We consider a three-level ladder atomic system as illustrated in Fig. \ref{fig:3-level-ladder}(c). Two successive transitions are coupled by a weak probe $\Omega_{p}$ and a strong control field $\Omega_{c}$, respectively. In the laboratory frame, the Hamiltonian for the atom-light interaction under the dipole approximation reads 
\begin{eqnarray}
\hat{H} = \sum^{N}_{j=1}\sum^{3}_{i=1}\hbar\omega_{i}\hat{\sigma}^{(j)}_{ii} - (\hat{\bm{d}}^{(j)}\cdot\bm{E}(\bm{r}_{j}) + H.c.)
\end{eqnarray}
where $\hbar\omega_{i}$ is the energy for the state $|i\rangle$, and $\hat{\bm{d}}^{(j)}$ is the dipole moment operator of the $j$-th atom. $N$ is the number of atoms, $H.c.$ stands for Hermitian conjugate. $\bm{E}(\bm{r}_{j})$ denotes the positive-frequency part of the electric field, which we can write as the sum of the two probe and control laser fields, i.e., $\bm{E}(\bm{r})=\bm{E}_{p}(\bm{r},t)+\bm{E}_{c}(\bm{r},t)=\bm{\mathcal{E}}_{p}(\bm{r})e^{i(\bm{k}_{p}\cdot\bm{r}-\omega_{p}t)} + \bm{\mathcal{E}}_{c}(\bm{r})e^{i(\bm{k}_{c}\cdot\bm{r}-\omega_{c}t)}$. Here $\bm{k}_{p},\omega_{p}$ and $\bm{k}_{c},\omega_{c}$ are the wave vectors and frequencies of the probe and control fields, respectively, and $\bm{\mathcal{E}}_{p}(\bm{r})$ and $\bm{\mathcal{E}}_{c}(\bm{r})$ denote the slowly-varying amplitude of each field. We further assume that the probe and control laser fields are near-resonant only with the respective atomic transitions, $|1\rangle\leftrightarrow|2\rangle$ and $|1\rangle\leftrightarrow|2\rangle$), such that the Hamiltonian simplifies to 
\begin{eqnarray}
\hat{H} &=& \sum^{N}_{j=1}\sum^{3}_{i=1}\hbar\omega_{ii}\hat{\sigma}^{(j)}_{ii} - [(\bm{d}_{21}\hat{\sigma}^{(j)}_{12} + \bm{d}_{12}\hat{\sigma}^{(j)}_{21})\cdot \bm{\mathcal{E}}_{p}(\bm{r}_{j})e^{i(\bm{k}_{p}\cdot\bm{r}-\omega_{p}t)}+ H.c.]\nonumber\\[2ex]
&&- [(\bm{d}_{32}\hat{\sigma}^{(j)}_{23} + \bm{d}_{23}\hat{\sigma}^{(j)}_{32})\cdot \bm{\mathcal{E}}_{c}(\bm{r}_{j})e^{i(\bm{k}_{p}\cdot\bm{r}-\omega_{p}t)} + H.c.]
\end{eqnarray}
where $\bm{d}^{(j)}_{12} = \langle 1_{j}|\hat{\bm{d}}^{(j)}|2_{j}\rangle=\bm{d}_{12}$ and $\bm{d}^{(j)}_{23} = \langle 2_{j}|\hat{\bm{d}}^{(j)}|3_{j}\rangle=\bm{d}_{23}$ are the dipole moments of the probe and control transition, respectively, and  $\hat{\sigma}^{(l)}_{ij}=|i_{l}\rangle\langle j_{l}|$ are the corresponding atomic transition operator for the $j$th atom. 

Transforming to a rotating frame via the unitary 
\begin{eqnarray}
\hat{U}=\exp\left[i\left(\sum_{j}\omega_{1}t\hat{\sigma}^{(j)}_{11}+(\omega_{1}+\omega_{p})t\hat{\sigma}^{(j)}_{22} + (\omega_{1}+\omega_{p} + \omega_{c})t\hat{\sigma}^{(j)}_{33}\right)\right],
\end{eqnarray}
and applying the rotating wave approximation, one obtains
\begin{eqnarray}
\frac{H}{\hbar}=-\sum_{j}(\Delta_{p}\hat{\sigma}^{j}_{22} + (\Delta_{p} + \Delta_{c})\hat{\sigma}^{(j)}_{33})
-\sum_{j}[\Omega_{p}(\bm{r}_{j})e^{i\bm{k}_{p}\cdot\bm{r}}\hat{\sigma}^{(j)}_{21}+\Omega_{c}(\bm{r}_{j})e^{i\bm{k}_{c}\cdot\bm{r}}\hat{\sigma}^{(j)}_{32} + H.c.]
\end{eqnarray}
where $\Delta_{p}=\omega_{p}-(\omega_{2}-\omega_{1})$ and $\Delta_{c}=\omega_{c}-(\omega_{3}-\omega_{2})$ are the frequency detunings of the probe and control field from their respective atomic transition, and $\Omega_{p}(\bm{r}_{j})=\bm{d}_{21}\cdot\bm{\mathcal{E}}_{p}(\bm{r})/\hbar$ and $\Omega_{c}(\bm{r}_{j})=\bm{d}_{32}\cdot\bm{\mathcal{E}}_{c}(\bm{r})/\hbar$ denote the Rabi frequencies of the two fields. For simplicity, we will assume that the beam profile of the control-field is significantly broader than that of the probe field, such that one can approximate  $\Omega_{c}(\bm{r}_{j})=\Omega_{c}$.

We are interested in the linear probe-field response of the atomic medium. Typically, the beam waist of the probe field is sufficiently large to describe its propagation dynamics by the paraxial wave equation 
\begin{eqnarray}
\label{eq:prop-eq1}
 \bigg[\frac{\partial}{\partial z} -\frac{i}{2k_{p}}(\frac{\partial^{2}}{\partial x^{2}} + \frac{\partial^{2}}{\partial y^{2}})\bigg]\mathcal{E}_{p}(\bm{r}) = i\frac{k_{p}}{\epsilon_{0}}{\bf e}^{*}_{p}\cdot \bm{P}(\bm{r})e^{-ik_{p}z}
\end{eqnarray}
where $\bm{\mathcal{E}}_{p}(\bm{r})= {\bf e}_{p}\mathcal{E}_{p}(\bm{r})$ and ${\bf e}_{p}$ is the unit polarization vector. Here, we assume a probe-field propagation along the $z$-direction, ${\bf e}_{z}$, such that the probe-field wave vector is $\bm{k}_{p}=k_{p}{\bf e}_{z}$. $\bm{P}(\bm{r})$ is the atomic polarization field $\bm{P}(\bm{r})=\sum_{j}\bm{d}_{21}\langle\hat{\sigma}^{(j)}_{12}\rangle\delta(\bm{r}-\bm{r}_{j})$, which is given by the atomic coherence $\langle\hat{\sigma}^{(j)}_{12}\rangle$. It is determined by the atomic dynamics that follows from the Heisenberg equations  
\begin{eqnarray}
    \frac{d\hat{\sigma}}{dt} = \frac{i}{\hbar}[\hat{H}_{\mathrm{int}},\hat{\sigma}] + \sum_{j}\frac{\Gamma_{2}}{2}[2\hat{\sigma}^{(j)}_{12}\hat{\sigma}\hat{\sigma}^{(j)}_{21} - \{\hat{\sigma}^{(j)}_{21}\hat{\sigma}^{(j)}_{12},\hat{\sigma}\}] + \frac{\Gamma_{3}}{2}[2\hat{\sigma}^{(j)}_{23}\hat{\sigma}\hat{\sigma}^{(j)}_{32} - \{\hat{\sigma}^{(j)}_{32}\hat{\sigma}^{(j)}_{23},\hat{\sigma}\}]
\end{eqnarray}
where $\Gamma_{2}$ and $\Gamma_{3}$ are the spontaneous decay rates of the excited states $|2\rangle$ and $|3\rangle$. Note that one can neglect Langevin noise terms in the above Heisenberg equations, when dealing only with normal ordered operators. The relevant atomic operators therefore evolve as
\begin{eqnarray}
    \frac{d\hat{\sigma}^{(j)}_{12}}{dt} &=& i(\Delta_{p}+i\frac{\Gamma_{2}}{2})\hat{\sigma}^{(j)}_{12} + i\Omega_{p}(\bm{r}_{j})e^{ik_{p}z_{j}}(\hat{\sigma}^{(j)}_{11}-\hat{\sigma}^{(j)}_{22}) + i\Omega_{c}e^{i\bm{k}_{c}\cdot\bm{r}_{j}}\hat{\sigma}^{(j)}_{13},\\[2ex]
    \frac{d\hat{\sigma}^{(j)}_{13}}{dt} &=& i(\Delta_{p}+\Delta_{c}+i\frac{\Gamma_{3}}{2})\hat{\sigma}^{(j)}_{13} + i\Omega^{*}_{c}e^{-i\bm{k}_{c}\cdot\bm{r}_{j}}\hat{\sigma}^{(j)}_{12}.
\end{eqnarray}

For $|\Omega_{c}|\gg |\Omega_{p}(\bm{r})|$, the atoms are only weakly excited, i.e., $\langle\hat{\sigma}^{(j)}_{11}\rangle=1$ and $\langle\hat{\sigma}^{(j)}_{22}\rangle=0$.This readily yields the steady state of $\langle\hat{\sigma}^{(j)}_{12}\rangle$ and allows one to reexpress the paraxial wave equation as
\begin{eqnarray}
\label{eq:prop-eq2}
 \bigg[\frac{\partial}{\partial z} -\frac{i}{2k_{p}}(\frac{\partial^{2}}{\partial x^{2}} + \frac{\partial^{2}}{\partial y^{2}})\bigg]\mathcal{E}_{p}(\bm{r}) = i\chi(\Delta_{p}) \mathcal{E}_{p}(\bm{r})
\end{eqnarray}
where the linear susceptibility $\chi(\Delta_{p})$ is given by
\begin{eqnarray}\label{eq:chi}
 \chi(\Delta_{p}) = -\frac{3n_{0}\lambda^{2}_{p}\Gamma_{2}}{8\pi}\frac{1}{\Delta_{p} + i\frac{\Gamma_{2}}{2} -\frac{|\Omega_{c}|^{2}}{\Delta_{p} +\Delta_{c} + i\frac{\Gamma_{3}}{2} }}.
\end{eqnarray}
in terms of the atomic density $n_{0}$ and the probe-field wavelength $\lambda_{p}=2\pi/k_{p}$. For $\Gamma_3=0$ and on two-photon resonance, $\Delta_p+\Delta_c=0$, the susceptibility   $\chi(\Delta_{p})$ vanishes, giving rise to EIT susceptibility~\cite{marangos2005}. Moreover, there are two absorption peaks that are readily understood in a dressed-state picture. To this end we can diagonalize the excited-state Hamiltonian 
\begin{eqnarray}
     \left[\begin{array}{cc}
       -\Delta_{p}  &  \Omega^{*}_{c}e^{-i\bm{k}_{c}\cdot\bm{r}}\\[2ex]
       \Omega_{c}e^{i\bm{k}_{c}\cdot\bm{r}}  & -\Delta_{p} -\Delta_{c}
    \end{array}\right]
\end{eqnarray}
that describes the coupling of $|2\rangle$ and $|3\rangle$ by the control field $\Omega_{c}$.  For $\Delta_{c}=0$, the  dressed eigenstates are simply $|\pm\rangle = (|2\rangle \pm |3\rangle)/\sqrt{2}$, with corresponding eigenfrequencies  $\lambda_{\pm}=\pm |\Omega_{c}|$. In between the two dressed-state energies, absorption is minimal are  two-photon resonance, $\Delta_{p}+\Delta_{c}=0$, and vanishes if $\Gamma_3=0$, as mentioned above.

Thermal motion affects the absorption primarily due to the Doppler shift of the atomic transitions. For a given atomic velocity $\bm{v}$, the dressed states are, thus, determined by 
\begin{eqnarray}
\left[\begin{array}{cc}
      -\Delta_{p}+k_{p}v_{z} & \Omega^{*}_{c}e^{-i\bm{k}_{c}\cdot\bm{r}}\\[2ex]
      \Omega_{c}e^{i\bm{k}_{c}\cdot\bm{r}} & -\Delta_{p}-\Delta_{c}+k_{p}v_{z}+\bm{k}_{c}\cdot\bm{v}
     \end{array}\right],
\end{eqnarray}
and have eigenfrequencies of 
\begin{eqnarray}
\label{eq:lambda-pm}
 \lambda_{\pm} = -\Delta_{p} + k_{p}v_{z} - \frac{1}{2}(\Delta_{c} - \bm{k}_{c}\cdot\bm{v}) \pm\frac{1}{2}\sqrt{(\Delta_{c} - \bm{k}_{c}\cdot\bm{v})^2 + 4|\Omega_{c}|^{2}}.
\end{eqnarray}

The second and third terms in Eq.~(\ref{eq:lambda-pm}) describe the single-photon Doppler effect on the transition frequency due to the atomic motion, while the last term introduces the frequency shift owing to the combined effect of the control coupling and the Doppler shift. 
The eigenvalues in Eq.~(\ref{eq:lambda-pm}) strongly depend on the relative propagation direction of the probe and the control field. Here, we consider two different configurations: (i) counter-propagating fields, such that $\bm{k}_{c}\cdot\bm{v}=-k_{c}v_{z}$, in terms of the velocity $v_z$ along the $z$-axis, and (ii) co-propagating fields such that $\bm{k}_{c}\cdot\bm{v}=k_{c}v_{z}$.

\begin{figure}[t!]
 \centering
 \includegraphics{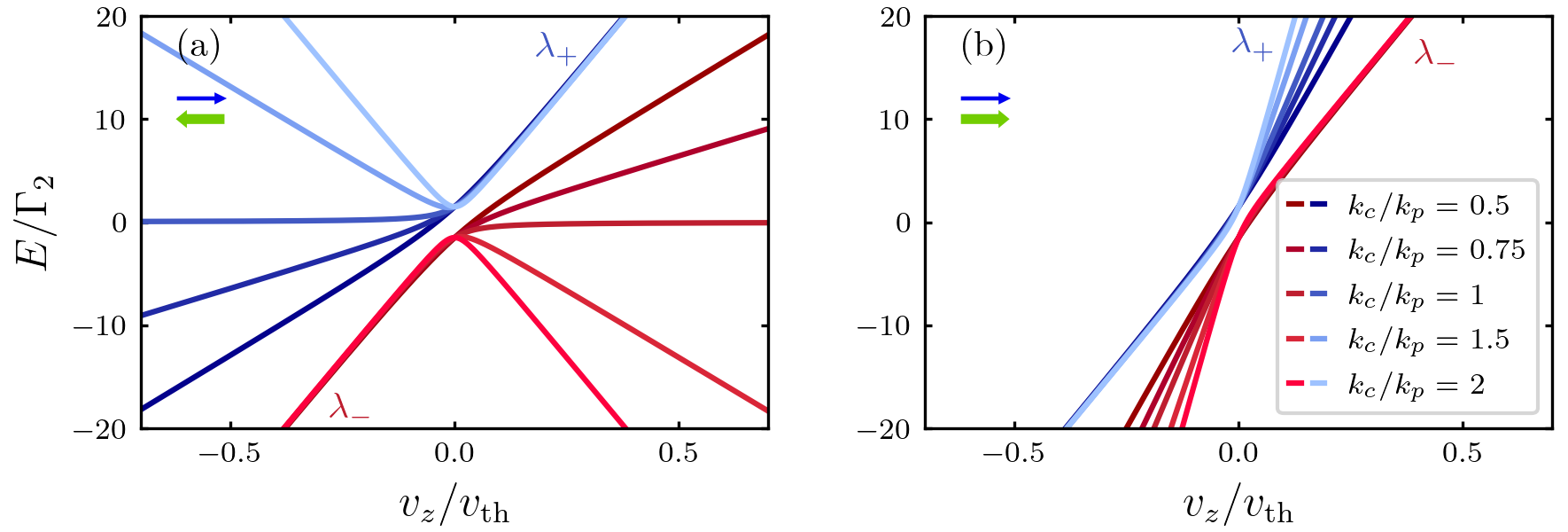} 
 \caption{(Color online) Comparison of eigenvalues $\lambda_{\pm}$ for different ratios $k_c/k_p$ for (a) counter- and (b) co-propagating probe and control fields in a ladder-type EIT scheme. For counter-propagating beams, a frequency gap opens up for $k_c>k_p$, whereas for the counter-propagating fields a frequency gap does not appear. Parameters are $\Delta_{c}=0,\Omega_{c}=1.5\Gamma_{2}$, $\Delta_{p}=0$.  $\Gamma_{2} = 2\pi\times 6.07\mbox{ MHz}, \Gamma_{3} = 2\pi\times 26.5\mbox{ kHz}$. The temperature is $T=320~K$.}
 \label{fig:co-and-counter-propagating}
\end{figure}

The characteristic velocity-dependence of the dressed-state eigenfrequencies is shown in Fig.~\ref{fig:co-and-counter-propagating}. 
For counter-propagating fields and large positive $v_{z}$, we have  $\lambda_{+}\simeq-\Delta_{p}+k_{p}v_{z}$ and $\lambda_{-}\simeq-\Delta_{p}-\Delta_{c}+(k_{p}-k_{c})v_{z}$. Hence, $\lambda_{+}$ always increases with $v_{z}$, while effect of $v_{z}$ on $\lambda_{-}$ depends on the sign of $k_{p}-k_{c}$. For $k_{c}/k_{p}>1$~(e.g., $k_{c}/k_{p}=1.5$ or $k_{c}/k_{p}=2.0$), the two states feature an energy gap as a function of $v_{z}$ [see Fig.~\ref{fig:co-and-counter-propagating}(a)]. If the frequency of the probe photon lies within this gap, absorbtion is greatly suppressed and a transparency window can emerge in the velocity-averaged absorption spectrum. For $k_{c}/k_{p}\leq 1$, this avoided crossing disappears and there will be no transparency window in the absorption spectrum.  

The eigenvalues for the co-propagating fields [cf. Fig. \ref{fig:3-level-ladder}(b)],  are shown in Fig.~\ref{fig:co-and-counter-propagating}(b). In this case,  no energy gap exists regardless of $k_{c}/k_{p}$, such that the transparency window is diminished by atomic motion. 

\begin{figure}[t!]
 \centering
 \includegraphics{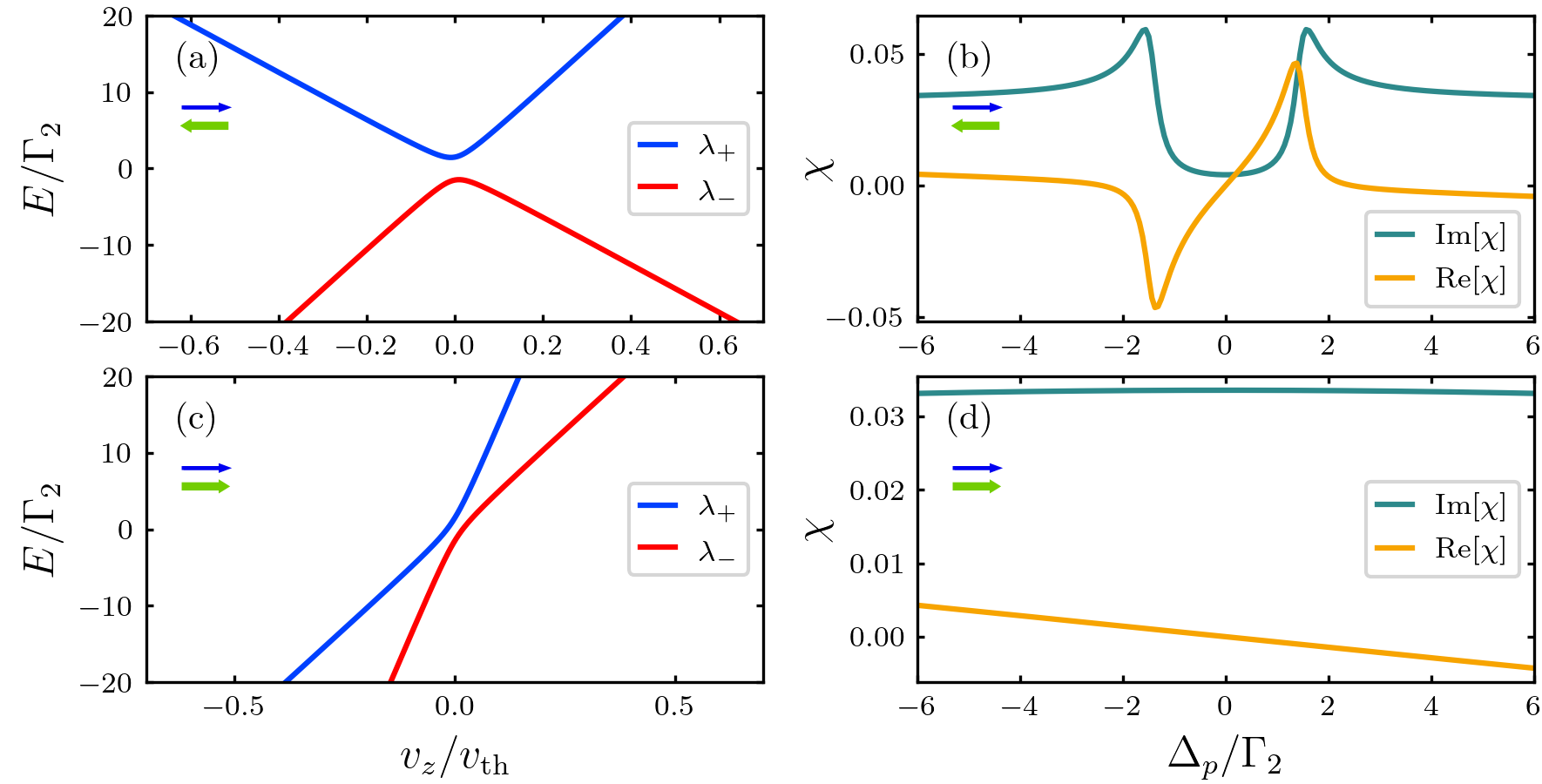} 
 \caption{(Color online) (a) The eigenvalues $\lambda_{\pm}$ as a function of $v_{z}$ exhibits a frequency gap when the two fields are counter-propagating~($\bm{k}_{p}=k_{p}\hat{\bm{e}}_{z},\bm{k}_{c}=-k_{c}\hat{\bm{e}}_{z}$),
 consequently, a box-shaped transparency window is created in the absorption spectrum plotted in (b). (c) The frequency gap is closed when the two fields are co-propagating~($\bm{k}_{p}=-k_{p}\hat{\bm{e}}_{z},\bm{k}_{c}=-k_{c}\hat{\bm{e}}_{z}$)
 leading to a Doppler-broadened absorptive spectrum shown in (d). Parameters are $\Delta_{c}=0,\Omega_{c}=1.5\Gamma_{2}$. In (a) and (c) we have taken $\Delta_{p}=0$. The three states are $5S_{1/2}, 5P_{3/2}$ and $21S_{1/2}$ of $^{87}\mathrm{Rb}$ with transition wavelengths to be $\lambda_{p}= 780.24~nm, \lambda_{c} = 488.08~nm$ and corresponding decay rates $\Gamma_{2} = 2\pi\times 6.07\mbox{ MHz}, \Gamma_{3} = 2\pi\times 26.5\mbox{ kHz}$. The temperature is $T=320~K$.}
 \label{fig:EigenEnergy-and-Spectrum}
\end{figure}

The absorption spectrum of a thermal gas is readily obtained from the average 
\begin{eqnarray}
 \chi(\Delta_{p}) =-\frac{3n_{0}\lambda^{2}_{p}\Gamma_{2}}{8\pi} \int^{\infty}_{-\infty}d\bm{v}\frac{f_{T}(\bm{v})}{\Delta_{p} - k_{p}v_{z} + i\frac{\Gamma_{2}}{2} -\frac{|\Omega_{c}|^{2}}{\Delta_{p}+ \Delta_{c} - k_{p}v_{z} -\bm{k}_{c}\cdot\bm{v} + i\frac{\Gamma_{3}}{2}}}
 \label{eq:Integrate_over_velocities}
\end{eqnarray}
of the optical susceptility [see Eq.(\ref{eq:chi})] over the Maxwell-Boltzmann distribution $f_{T}(\bm{v})=(\pi v^{2}_{\mathrm{th}})^{-3/2}e^{-v^{2}/v^{2}_{\mathrm{th}}}$ of the atomic velocities. Here, $v_{\mathrm{th}} = \sqrt{2k_{B}T/m}$ is the most probable velocity, and $T$ and $m$ are the temperature and mass of the atoms, respectively. The characteristic spectrum of the susceptibility is shown in Fig.~\ref{fig:EigenEnergy-and-Spectrum} for the two field configurations along with the velocity dependence of the dressed-state energies. Indeed, there is a broad transparency window for counter-propagating fields, that can be broader than the EIT window for stationary atoms. Outside of the transparency window there is a broad absorption background due to the Doppler broadening of the probe-field resonance. In stark contrast, one finds  flat feature-less absorption background in the co-propagating case. Transparency, thus, emerges from a \emph{negative wavenumber mismatch}, which can not be satisfied for co-propagating fields.

\begin{figure}[t!]
 \centering
 \includegraphics{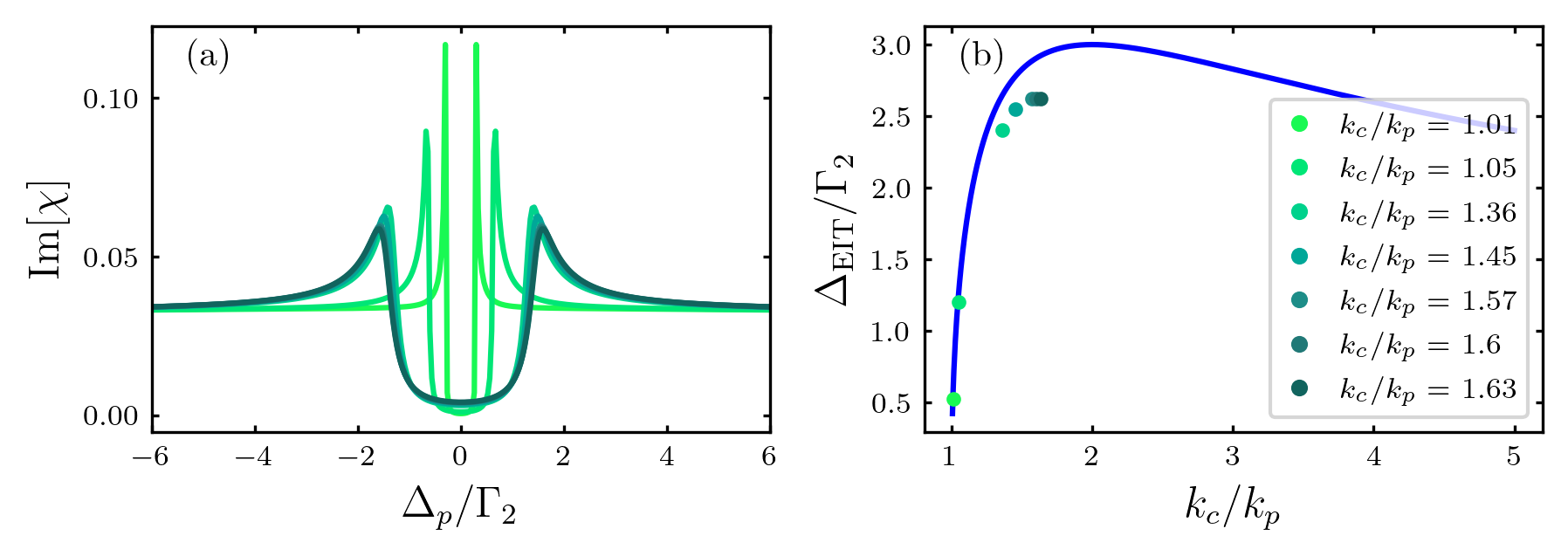} 
 \caption{(Color online) (a) The absorption spectrum for different ratio $k_{c}/k_{p}$. Here $k_{c}$ is changed to show its effect on the transparency window, the ratio $k_{c}/k_{p}$ in increasing order corresponds to the case of state 3 chosen to be $5D_{5/2}, 7S_{1/2}, 7D_{5/2}, 10S_{1/2}, 15S_{1/2}, 20S_{1/2}, 100S_{1/2}$ respectively. (b) The solid line shows the approximate width of the EIT window defined by Eq.~(\ref{eq:EITWidth}), where the points are obtained from fitting the EIT spectrum plotted in (a). The fitting width is defined as the detuning range in which the absorption is smaller than the background in the two wings. Parameters are the same as in Fig.~\ref{fig:EigenEnergy-and-Spectrum}.}
 \label{fig:EIT-spectru-vs-kratio}
\end{figure}

To further illustrate this effect, we calculate the EIT spectrum for different ratios of $k_{c}/k_{p}$ for counter-propagating fields. Fig.~\ref{fig:EIT-spectru-vs-kratio} illustrates the emergence of a transparency window  for $k_{c}/k_{p}\geq1$. The width of the transparency window depends on the frequency gap, which is approximately given by
\begin{eqnarray}
\label{eq:EITWidth}
\Delta_{\mathrm{EIT}} \simeq \frac{\sqrt{k_{p}(k_{c}-k_{p})}}{k_{c}}\cdot 4\Omega_{c}
\end{eqnarray}
for $\Delta_{p}=\Delta_{c}=0$.
For $k_{c}=k_{p}$ one still observes a narrow EIT window, since in this case the energies of the two dressed states only coincide asymptotically for $v_z\rightarrow\infty$. 

 Eq.~(\ref{eq:EITWidth}) does not account for the finite linewidths of the dressed states, which are given by 
\begin{eqnarray}
\Gamma_{+}  = \Gamma_{-}  = \left[1-\frac{k_{p}}{k_{c}}\right]\Gamma_{2}
\end{eqnarray}
at the state-crossing. 
When $k_{c}/k_{p}\simeq 1$, $\Gamma_{\pm}$  is very small such that Eq.~(\ref{eq:EITWidth}) agrees well with the actual width of the EIT spectrum [see  Fig.~\ref{fig:EIT-spectru-vs-kratio}(b)]. For larger values of $k_{c}/k_{p}$, however, the finite linewidth of both states causes deviations from Eq.~(\ref{eq:EITWidth}) and slightly reduces the effective width of the EIT window as shown in Fig.~\ref{fig:EIT-spectru-vs-kratio}(b). 
Nevertheless, the EIT window in hot atoms can be broader than that in cold atoms for a given set of parameters. In this regard, the described negative wavenumber mismatch does not only recover EIT but also improves the transmission of in hot atoms. We finally note that the linear scaling with $\Omega_{c}$ [cf. Eq.~(\ref{eq:EITWidth})] implies that even comparatively weak control fields can generate a significant EIT effect. This is in contrast to a recent experiment on hot atoms~\cite{nori2021} in which the generation of transparency required large control-field intensities. This can be understood from the above discussion since the experiment used a state configuration for which $k_{c}<k_{p}$ and does therefore not exploit the effect of wavenumber mismatch described above.  

\section{Experimental demonstration}
\begin{figure}[t]
 \centering
 \includegraphics{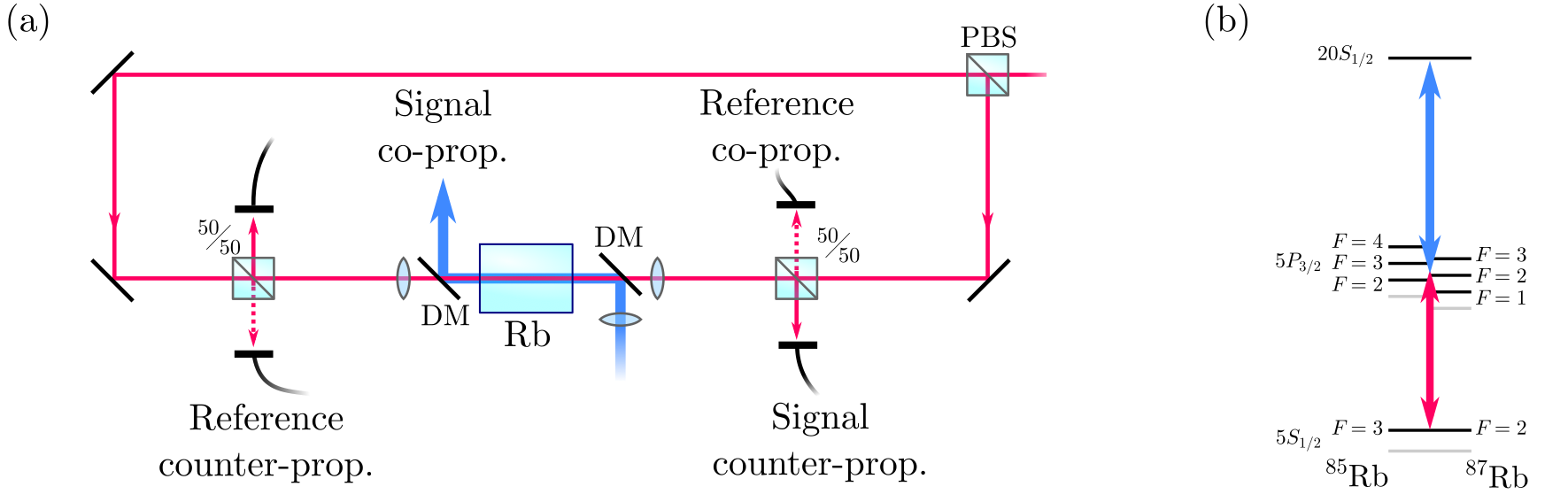} 
 \caption{(Color online) Schematics of experimental setup for measuring hot EIT with probe and control beams in co- and counter-propagating configuration. The probe beams for the two configurations are split from the same incoming beam and passed through the setup in opposite direction.
 Before the vapor cell, each beam is split on a beam splitter to obtain a reference signal without atoms.
 The transmission signal through the vapor cell and the reference signal are fiber coupled and measured on single-photon counter modules. The resulting transmissions are shown in Fig.~\ref{Theory-and-Experiment}. The control beam is overlapped with the probe beam on two dichroic mirrors (DM).}
 \label{Experimental-setup}
\end{figure}

To demonstrate the EIT recovery with $k_c>k_p$ we perform a proof of principle experiment in a room-temperature vapor of Rb atoms. The setup is shown in Fig.~\ref{Experimental-setup} (b) along with the level scheme of the transitions used for the probe and control light in Fig.~\ref{Experimental-setup} (b). The experiments are performed in a glass cell with length $L = 5\,\mathrm{cm}$ that contains the isotopes $^{85}$Rb and $^{87}$Rb at their natural abundance and at $T\approx 296\,\mathrm{K}$.

The probe light with wavelength $\lambda_p=780\,\mathrm{nm}$ couples the ground state $|1\rangle = |5S_{1/2},F=2\rangle$ to a hyperfine manifold of intermediate states $|2\rangle = |5P_{3/2},F=1,2,3\rangle$ in $^{87}$Rb and $|1\rangle = |5S_{1/2},F=3\rangle$ to $|2\rangle = |5P_{3/2},F=2,3,4\rangle$ in $^{85}$Rb. 
We focus the probe light to a waist of $\approx 50\,\mathrm{\mu m}$.
The control light with wavelength $\lambda_c= 488\,\mathrm{nm}$ and intensity $0.66 \, \mathrm{W}$ couples $|2\rangle$ to a low-lying Rydberg state $|3\rangle = |20S_{1/2}\rangle$ and is focused to a waist of $\approx 70\,\mathrm{\mu m}$.

The peak Rabi frequency in the focus is $\approx 19 \, \Gamma $, while the effective Rabi frequency across the entire length of the vapor cell is significantly lower. The discrepancy in wavelength leads to $k_c/k_p=\lambda_p/\lambda_c\approx 1.6$. Fig.~\ref{fig:EIT-spectru-vs-kratio} shows that one expects an EIT window with significant width for this ratio. 

To investigate both co- and counter-propagating probe and control fields, the setup is build symmetrically, such that the probe light can be sent through the glass cell from both sides.
After the cell, the probe light is coupled into a single-mode fiber and detected on a single-photon counter. This signal can be compared to a reference signal picked up before the glass cell and detected on a different single-photon counter.

\begin{figure}[t!]
 \centering
 \includegraphics{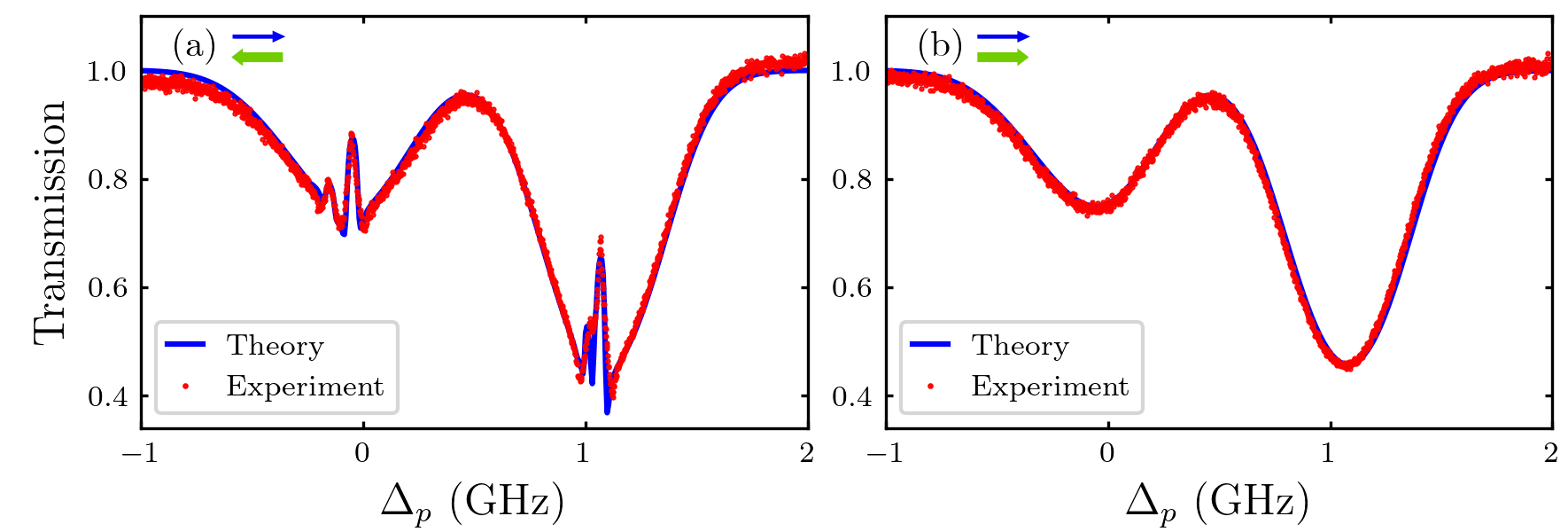} 
 \caption{(Color online) 
 Experimental observation of direction-depndent recovery of EIT in a Rb vapor for counter= (a)  and co-propagating (b) probe and control beams. Besides the experimental data (red) we also show the theoretically predicted transmission (blue) for $\Delta_{c} = 13\Gamma_{2}, \Omega_{c} = 7\Gamma_{2}$ and additional phenomenological decay from the intermediate state with rates $\gamma_{87}=5\Gamma_{2}$ and $\gamma_{85}=3\Gamma_{2}$ for $^{87}\mathrm{Rb}$ and $^{85}\mathrm{Rb}$ respectively. All other parameters are the same as in Fig.~\ref{fig:EigenEnergy-and-Spectrum}.}
 \label{Theory-and-Experiment}
\end{figure}

Fig.~\ref{Theory-and-Experiment} (a) and (b) show the transmission of a weak probe beam with peak intensity a factor of $> 100$ below saturation as we scan the probe detuning $\Delta_p$ over the Doppler-broadened probe transitions for both isotopes.
For counter-propagating probe and control light, we observe multiple EIT resonances that arise from the different hyperfine levels of the intermediate state $|2\rangle$.  In the case of copropagating probe and control, the EIT resonances are absent as predicted by theory.
Comparing the experimental data to theory, we find excellent agreement of the respective transmission curves for $\Delta_{c} = 13\Gamma_{2}, \Omega_{c} = 7\Gamma_{2}$. 

Besides spontaneous decay from the intermediate state $|2\rangle$ with rate $\Gamma_2$, we have also phenomenologically introduced additional decay from $|2\rangle$ with rates $\gamma_{87}=5\Gamma_{2}$ (for $^{87}\mathrm{Rb}$) and $\gamma_{85}=3\Gamma_{2}$ ($^{85}\mathrm{Rb}$). This additional decay would include the diffraction of the probe beam which causes absorption when its size is larger than the control beam, imperfect polarization of the laser beams, collisional broadening and other experimental imperfections \cite{Singh2009,Firstenberg2023}.

\section{Summary and outlook}
In this tutorial we have presented a theoretical analysis of a Doppler-broadened three-level ladder-system and pointed out that while atomic motion limits the visibility of EIT for certain positive wavenuber mismatch,  EIT is recovered and enhanced for negative wavenumber mismatch where the control field has a wavenumber larger than that of the probe field.

We have demonstrated this effect in an experiment with hot Rubidium atoms where a Rydberg state is used to obtain a wavevector ratio so that $k_p<k_c$ is satisfied for counter-propagating probe and control fields. In the experiment, the width of the EIT window is limited by available control beam power, but the calculations show that it scales favorably with the wavelength mismatch in hot atomic vapors and can even exceed what can be reached without Doppler broadening for the same laser powers. 

The mechanism discussed here is not limited to the three-level ladder-system. Similar conditions on the wavenumber mismatch exists for lambda systems \cite{dunn1999}, and the theory may be further extended to systems with more levels and coupling lasers to employ negative three-photon or even high-order wavenumber mismatch \cite{Singh2012,Raithel2019}. A direct application of the non-reciprocal transmission in the three-level system is the realization of magnetic field-free high-fidelity optical isolators \cite{Xiao2018,Gong2019,Shi2020,nori2021}. The relative simplicity of the scheme lends itself to combination with nano-optical structures or miniaturized vapor cells to realize on-chip optical isolators and circulators \cite{Gaeta2005,Russell2005,Pfau2010,Adams2012b,Rauschenbeutel2015,Rauschenbeutel2016,Levy2021,Treutlein2023}. Finally, as for alkali atoms $k_c>k_p$ is satisfied when a ladder-system with a Rydberg state is used, this opens the opportunity of combining enhanced EIT with Rydberg-mediated photon-photon interaction for realization of nonlinear quantum optics \cite{Adams2013c,Lukin2014b,Pohl2016d,Hofferberth2016d,Pfau2018}. 

All of these points serve to highlight how the Doppler effect in hot atomic vapors, which is often seen as a hindrance, can be utilized in EIT systems to realize nonreciprocal optics in hot atoms with wide-ranging applications.

\section*{Acknowledgements}
This work was supported by the European Union through the Horizon 2020 program under the ERC consolidator grant RYD-QNLO (grant no. 771417) and through the Horizon Europe ERC synergy grant SuperWave (grant no. 101071882). The work was also supported by the Carlsberg Foundation through the Semper Ardens Research Project QCooL.

\section*{Data availability}
The experimental data for Fig.~\ref{Theory-and-Experiment}, and the code for creating all figures in this tutorial are available in the Zenodo database under accession code \href{https://doi.org/10.5281/zenodo.10689011}{https://doi.org/10.5281/zenodo.10689011}

\section*{References}
\bibliographystyle{iopart-num}
\bibliography{references}

\end{document}